\documentclass[a4paper,fleqn,usenatbib]{mnras}

\usepackage[T1]{fontenc}
\usepackage{aecompl}%{ae,aecompl}

\usepackage[utf8]{inputenc}
\usepackage{bm}

\usepackage{graphicx}	% Including figure files
\usepackage{amsmath}	% Advanced maths commands
\usepackage{amssymb}	% Extra maths symbols
\usepackage{xcolor}
\usepackage{ulem}
\newcommand{\bh}{{\rm{BH}}}
\newcommand{\disc}{{\rm{disc}}}
\newcommand{\infl}{{\rm{in}}}
\newcommand{\fedd}{f_{\rm{Edd}}}

\newcommand{\isco}{{\rm{ISCO}}}

\newcommand{\warp}{{\rm{warp}}}
\newcommand{\sg}{{\rm{sg}}}

\newcommand{\msun}{~\text{\rm{M}$_{\sun{}}$}}
\newcommand{\pc}{~\rm{pc}}

\title[Black hole spin evolution in warped accretion discs]{Black hole spin evolution in warped accretion discs}

\author[E. Cenci et al.]{Elia Cenci,$^{1}$\thanks{E-mail: e.cenci@campus.unimib.it}
Luca Sala,$^{1,2}$
Alessandro Lupi,$^{3}$
Pedro~R. Capelo$^{4}$
and Massimo Dotti$^{1,5}$
\\
$^{1}$Dipartimento di Fisica ``G.~Occhialini'', Universit\`{a} degli Studi di Milano-Bicocca, Piazza della Scienza 3, IT-20126 Milano, Italy\\
$^{2}$ Universitäts-Sternwarte München,Fakultät für Physik, LMU Munich,  Scheinerstr. 1, D-81679 München, Germany
$^{3}$Scuola Normale Superiore, Piazza dei Cavalieri 7, IT-56126 Pisa, Italy\\
$^{4}$Center for Theoretical Astrophysics and Cosmology, Institute for Computational Science, University of Zurich, Winterthurerstrasse\\ 190, CH-8057 Z\"urich, Switzerland\\
$^{5}$INFN, Sezione di Milano-Bicocca, Piazza della Scienza 3, IT-20126 Milano, Italy
}

\date{Accepted 2020 October 30. Received 2020 October 14; in original form 2020 July 15}

\pubyear{2020}

\begin{document}
\label{firstpage}
\pagerange{\pageref{firstpage}--\pageref{lastpage}}
\maketitle

\begin{abstract}
Massive black holes (BHs) inhabiting galactic nuclei can be described by two parameters only, i.e. mass and spin, that change through cosmic time in response to accretion and merger events. While most numerical simulations accurately track the BH mass, spin evolution is rarely taken into account. In this work, we implement and validate a self-consistent sub-grid model for the evolution of the BH mass and spin via gas accretion in the hydrodynamics code {\textsc{gizmo}}. The model assumes that accretion from resolved scales does not occur instantaneously, but is mediated by a sub-grid geometrically thin $\alpha$-disc. After validating our model semi-analytically, we test it in an idealised environment consisting of a circumnuclear disc, where gas accretion on to the accretion disc is consistently determined by \textsc{gizmo}. In the absence of any accretion-related feedback, the spin evolution closely traces that observed in the semi-analytical models, and depends on the free parameters of our implementation, such as the initial BH spin, angular momentum of the accretion disc, and radius at which the gas inflow circularises. In \textsc{gizmo}, we also couple our model with the biconical-outflow model presented in a companion paper, wherein the feedback axis is always aligned with the BH spin. In this last case, the evolution of the central BH differs significantly from the previous cases, since the feedback process modifies the gas dynamics and its inflow rates from resolved scales. Such an interaction cannot be modelled by simple semi-analytical models and should be treated using full $N$-body hydrodynamical simulations.
\end{abstract}


%%%%%%%%%%%%%%%%%%%%%%%%%%%%%%%%%%%%%%%%%%%%%%%%%%

%%%%%%%%%%%%%%%%% BODY OF PAPER %%%%%%%%%%%%%%%%%%

%%%%%%%%%%%%%%%%%%%%%%%%%%%%%%
%%%%%%%%%%%%%%%%%%%%%%%%%%%%%%
%%%Introduction
%%%%%%%%%%%%%%%%%%%%%%%%%%%%%%
%%%%%%%%%%%%%%%%%%%%%%%%%%%%%%

\section{Introduction}

According to the no-hair conjecture \citep{Israel67, Israel68, Carter71, Hawking72, Robinson75}, massive black holes (BHs) are thought to be completely characterized by three parameters: mass, charge, and spin. Since any electric charge would be quickly neutralized by charges in the surrounding plasma, astrophysical BHs can be described only in terms of their masses and spins.

Spins in particular have a number of fundamental consequences on the evolution of the BHs. The spin magnitude affects the position of the innermost stable circular orbit \citep[ISCO;][]{B1970, B1972} and, as a consequence, the radiative efficiency of accretion processes\footnote{The dependence of the radiative efficiency on the spin is weaker for supercritically accreting BHs, see, e.g. \citet{MHD14}.} and the rate at which BHs can grow in time. Furthermore, according to the spin paradigm, high spins are responsible for the launching of the relativistic jets observed in active galactic nuclei (AGN) over the whole spectral range \citep{BZ1977}. 
 
The spin direction as well has been proposed to play a central role in the growth of the BHs and in the co-evolution with their host galaxies. As an example, feedback exerted by the accreting BH on to its environment has been suggested as a possible cause for the observed BH-host galaxy scaling relations \citep[][]{Kormendy_Ho_2013} and for the quenching of star formation in massive galaxies \citep{SR98, Fabian99, Fabian12}. It has been argued that, to achieve such effects, the spin direction\footnote{The feedback reference direction is expected to be parallel to the BH spin both for a \citet{BZ1977} jet as well as for a wind launched from the very central regions of an accretion disc \citep{BP1975}.} must evolve in time (see, e.g. \citealt{NPK12} for the galaxy-scale feedback from a wind and \citealt{Cielo18} for the feedback on to the intra-cluster medium mediated by a jet).

Finally, spin magnitudes and directions determine the recoil velocity that the remnant of a BH merger experiences due to anisotropic gravitational-wave emission \citep[e.g.][]{Kop2007}, with such velocity being well above typical escape velocities from massive galaxies for some spin configurations \citep[e.g.][]{Campanelli07, Baker08, Herrmann07, Schnittman07, LZ11, Lousto12}.

The spin direction has a direct impact on the evolution of the spin magnitude. Prolonged accretion events with a fixed geometry result in maximally spinning BHs after a  mass growth comparable to the initial BH mass \citep{B1970}. On the contrary, episodic accretion events isotropically oriented and each having an accreted mass significantly smaller than the initial mass of the BH will on average decrease the BH spin, due to the larger size of the ISCO and therefore a larger magnitude of the  angular momentum per unit of mass associated to retrograde accretion events \citep{KP06}. 

\cite{D2013} showed that the two above-mentioned accretion modes, often dubbed ``coherent'' and ``chaotic'', are the extremes of all the possible accretion configurations, and that, depending on the assumptions on the BH fueling geometry, the expected spin magnitudes for any given BH mass can seamlessly vary from 0 to $\sim$1, in agreement with the results of early $N$-body hydrodynamical simulations with the spin of the central BH being evolved in post-processing \citep[][]{Dotti_et_al_2010,Maio_et_al_2013} or on-the-fly \citep[][]{Dubois_et_al_2014b,Dubois_et_al_2014a}. By implementing the model by \cite{D2013} in a pre-existing semi-analytical galaxy formation model \citep{Barausse12}, and by assuming that the gas reservoir for the BH growth has the same dynamical properties of the host galaxy nuclei at $\gtrsim 100$~pc scales,\footnote{Such assumption has still to be proven correct, via statistically significant samples of both high-resolution observations and simulations.} \cite{S2014} managed for the first time to reproduce the observational constraints on BH spins available, without the introduction of any additional freely-tunable parameter.

The above-cited studies about accretion-driven spin evolution did not consider the possible effect of AGN feedback on to the dynamics of the gas fueling the accretion process. As an example, a prolonged accretion event whose accretion disc is initially misaligned with respect to the BH spin direction by more than $\pi/2$ would tend to re-align the BH spin with the gas angular momentum \citep[][]{BP1975}. If, however, the accretion process triggers a directional feedback aligned with the BH spin, the feedback could significantly alter the dynamics of the gas reservoir, modifying the following spin evolution in a non-linear fashion. This is a severe limitation of semi-analytical studies, that cannot follow in real time the impact of the spin evolution on the larger-scale gas dynamics. For this reason, we hereby present a new implementation for the  coupled evolution of BH spins, unresolved accretion discs, and directional feedback in $N$-body, hydrodynamical simulations. Our implementation includes the spin evolution discussed in \cite{F2018}, that relaxes the small-warp approximation  \citep{SF1996, M2007, P2009, D2013}, the directional feedback presented in a companion paper \citep{Sala2021}, and a new sub-grid model for the self-consistent  evolution of unresolved accretion discs around BHs. The model has been implemented in the publicly available code {\textsc{gizmo}}\footnote{\url{http://www.tapir.caltech.edu/~phopkins/Site/GIZMO.html}} \citep{H2015}.

The manuscript is structured as follows: the new model for the unresolved accretion disc and spin evolution is described in Section~\ref{sec:spin_evolution_model}, and its semi-analytical validation (in the absence of feedback) is presented in Section~\ref{sec:validation}. The initial conditions of the idealized run used to test our full implementation are discussed in Section~\ref{sec:setup}. The results of the tests and their discussion are presented in Sections~\ref{sec:runs} and \ref{sec:conclusions}, respectively.

%%%%%%%%%%%%%%%%%%%%%%%%%%%%%%
%%%%%%%%%%%%%%%%%%%%%%%%%%%%%%
%%%The spin evolution model
%%%%%%%%%%%%%%%%%%%%%%%%%%%%%%
%%%%%%%%%%%%%%%%%%%%%%%%%%%%%%

\section{The spin evolution model}\label{sec:spin_evolution_model}

In this section, we introduce our spin evolution model and validate it in a semi-analytic controlled environment. In our model, a BH particle represents a sub-grid system composed of a BH surrounded by a standard $\alpha$-disc \citep[][]{SS1973}.

%%%%%%%%%%%%%%%%%%%%%%%%%%%%%%
%%%%%%%%%%%%%%%%%%%%%%%%%%%%%%
%%%Model description
%%%%%%%%%%%%%%%%%%%%%%%%%%%%%%
%%%%%%%%%%%%%%%%%%%%%%%%%%%%%%

\subsection{Model description}\label{sec:model}

The BH is described by its mass $M_\bh = 10^6 M_{\bh,6} \msun$, angular momentum direction $\bm{j}_\bh = \bm{J}_\bh/J_\bh$, and spin parameter\footnote{In this work, we define the spin parameter as a dimensionless quantity varying between 0 for a Schwarzschild BH and 1 for an isolated, maximally spinning BH, as opposed to other works where $a_\bh = J_\bh/(M_\bh c)$, with the maximal spin being $GM_\bh /c^2$.} $a_\bh = c J_\bh /(G M_\bh^2)$, where $\bm{J}_\bh$ and $J_\bh$ are the BH's angular momentum vector and magnitude, respectively, $c$ is the speed of light in vacuum, and $G$ is the gravitational constant.

To characterize the accretion disc, we need to specify its mass $M_\disc = 10^4 M_{\disc ,4} \msun$, total angular momentum $\bm{J}_\disc = \bm{j}_\disc J_\disc$ (where $\bm j_\disc$ and $J_\disc$ are the angular momentum direction and magnitude, respectively), and the accretion rate on to the BH $\dot{M}_{\rm acc-BH}$. The Eddington ratio can be expressed by $\fedd = \dot{M}_{\rm acc-BH} / \dot{M}_{\rm Edd}$, where $\dot{M}_{\rm Edd} = 4\pi G M_\bh m_{\rm p}/(\sigma_{\rm T}\eta c)$ is the Eddington accretion rate, $m_{\rm p}$ is the proton mass, $\eta = 0.1\eta_{0.1}$ is the radiative efficiency, and $\sigma_{\rm T}$ is the Thomson cross-section. In order to initialise the accretion disc properties, we set $\bm{j}_\disc=\bm{j}_{\rm{gas}}$, where $\bm{j}_{\rm{gas}}$ is the angular momentum direction of the gas surrounding the BH-disc system.

Following the \citet{SS1973} solution for the external regions of the disc, the radial viscosity $\nu_1$  scales with the cylindrical radius $R$ as \citep[][]{M2007}

\begin{equation}
	\nu_1\,=\,A_{\nu_1}\,R^\beta
	,
	\label{eqn:nu_1}
\end{equation}

\noindent where $\beta = 3/4$. The normalization is computed as \citep[][]{F2002,P2009}

\begin{equation}
	A_{\nu_1}\,\simeq\,9\times10^6\,\alpha_{0.1}^{4/5}\,M_{\bh,6}^{1/20}\,\left(\frac{\fedd}{\eta_{0.1}}\right)^{3/10}
	~\rm{cm}^{5/4}~\rm{s}^{-1} 
	,
	\label{eqn:A_nu1}
\end{equation}

\noindent where $\alpha = 0.1 \alpha_{0.1}$ is related to the definition of $\nu_1$ (via $\nu_1 \equiv \alpha c_{\rm s} H$), $c_{\rm s}$ is the gas sound speed, and $H$ is the disc scale-height. We compute the vertical viscosity $\nu_2$ under the approximation of a constant viscosity ratio $\nu_2/\nu_1 = \xi\alpha^{-2}/2$ \citep[][]{LP2007}. In order to ensure that the adopted analytical prescription for viscosity is valid, we assume that vertical perturbations in the disc propagate diffusively, the latter translating in $\alpha > H/R$ \citep{P1992}. For our tests, we set $\alpha = 0.1$ inside the disc, in agreement with the value inferred by available observational data for thin accretion discs \citep[][]{K2007}, and  $\xi = 0.7$ \citep[][]{LP2007,P2009}. 
By equating the Lense--Thirring \citep[][]{LT1918} precession time-scale $\Omega_{\rm LT}^{-1} = c^2 R^3 / (2 G J_\bh)$ to the characteristic time-scale $\tau_{\nu_2} \sim R^2 / \nu_2$ of the vertical perturbation propagation in the disc out to a radius $R$ \citep[][]{LP2006}, we obtain the characteristic extension of the warp $R_\warp$, that divides the disc in two regions: the inner one ($R \ll R_\warp$), where the disc and BH angular momenta are expected to be parallel, and the outer one ($R \gg R_\warp$), where the disc maintains its original inclination ($\bm{j}_\disc \sim \bm{j}_{\rm gas}$). For the adopted viscosity model, we obtain

\begin{equation}
	\frac{R_\warp}{R_{\rm g}}\,\simeq\,952\,\xi^{-4/7}\,M_{\bh,6}^{4/35}\,\left(\frac{\fedd}{\eta_{0.1}}\right)^{-6/35}\,a_\bh^{4/7} 
	,
\end{equation}

\noindent where $R_{\rm g} = G M_\bh / c^2$ is the BH gravitational scale-radius. The radiative efficiency $\eta$ in the disc is determined self-consistently from the properties at the equatorial ISCO of a \citet{Novikov_Thorne_1973} disc \citep[][]{B1970,B1972}, under the assumption that the inner part of the disc is aligned with the BH equatorial plane due to the Bardeen--Petterson effect \citep[][]{BP1975}:

\begin{equation}
	\eta\,=\,1\,-\,\sqrt{1\,-\,\frac{2}{3}\frac{R_{\rm g}}{R_\isco}\;}
	,
	\label{eqn:eta}
\end{equation}  

\noindent where $R_\isco$ is the ISCO radius, computed as in \citet[]{B1972}. Assigning a viscosity prescription allows to compute the steady-state solution for the disc mass and angular momentum density profiles (i.e. per unit surface). The disc total angular momentum is defined as the integrated angular momentum density out to an orbit with $R = R_{\rm out}$ that encompasses a mass equal to the disc total mass $M_\disc$. Assuming that the main contribution to the disc total angular momentum comes from the outer regions of the disc, we can consistently derive the total angular momentum magnitude as \citep[][]{F2018}

\begin{equation}
	\frac{J_\disc}{J_\bh}\,\simeq\,2.8\,\alpha_{0.1}^{8/25}\,M_{\bh,6}^{-47/25}\,M_{\disc ,4}^{7/5}\,\left(\frac{\fedd}{\eta_{0.1}}\right)^{-7/25}\,a_\bh^{-1} 
	.
	\label{eqn:J_disc/J_bh}
\end{equation}

Once the BH-disc system is initialised, the time evolution is modelled according to the prescriptions of \citet{F2018}, consistently with the properties of the large-scale resolved environment. At every time-step $\Delta t$, we compute the Eddington ratio $\fedd$ by means of Equation~\eqref{eqn:J_disc/J_bh}. To ensure sub-Eddington accretion, as required by the assumed disc model, we enforce $\fedd \le 1$. The standard thin-disc model is not supposed to hold in low-accretion regimes. In those cases, one should instead implement a more appropriate disc model (e.g. an advection-dominated accretion flow; \citealt{Narayan_Yi_1994}). However, we do not impose any lower limit to $\fedd$ \citep[as done in, e.g.][]{F2018}, because the validity threshold is not well constrained \citep[see, e.g.][]{Chen_et_al_1995} and the BH spin evolution is negligible for very small accretion rates.

The BH and disc masses are updated at every time-step taking into account the disc draining rate $\dot{M}_{\rm acc-BH}$, radiative losses (through $\eta$), the mass accretion rate on to the disc $\dot{M}_\infl$ given by the dynamics of the resolved-scales gas, and the mass outflow rate $\dot{M}_{\rm out}$:

\begin{align}
	\dot{M}_\bh\,&=\,\left(1-\eta\right)\,\dot{M}_{\rm acc-BH} , \\
	\dot{M}_\disc\,&=\,\dot{M}_\infl\,-\,\dot{M}_{\rm acc-BH}\,-\,\dot{M}_{\rm out} 
	.
	\label{eqn:dM/dt}
\end{align}

The BH angular momentum evolves in response to both the gas accretion at $R = R_\isco$ and the gravito-magnetic torque exerted by the material flowing through $R \sim R_\warp$. While any process taking place at the ISCO only modifies the magnitude of the BH angular momentum, the Bardeen--Petterson torque, in steady-state warped conditions, is responsible for its change in direction \citep[][]{K2005,F2018}:

\begin{align}
	\dot{\bm{J}}_\bh\,&=\,\dot{M}_{\rm acc-BH}\,\Lambda_\isco\,\frac{\bm{j}_\bh\cdot\bm{j}_\disc}{|\bm{j}_\bh\cdot\bm{j}_\disc|}\,\bm{j}_\bh\nonumber\\
	&-\,\frac{\bm{J}_\bh}{\tau_{\rm{gm}}}\times\left\{\sin\left(\frac{\pi}{7}\right)\bm{j}_\disc\,+\,\cos\left(\frac{\pi}{7}\right)\left(\bm{j}_\bh\times\bm{j}_\disc\right)\right\} 
	.
	\label{eqn:dJ_bh/dt}
\end{align}

Here $\Lambda_\isco$ is the specific angular momentum per unit mass carried by the gas flowing at the ISCO, and $\tau_{\rm{gm}}$ is the characteristic time-scale over which the gravito-magnetic interaction significantly changes the BH spin direction \citep[][]{M2007,LP2006,P2009,D2013}:

\begin{equation}
	\tau_{\rm{gm}}\,\simeq\,0.17\,\xi^{-5/7}\,\alpha_{0.1}^{58/35}\,M_{\bh,6}^{-2/35}\,\left(\frac{\fedd}{\eta_{0.1}}\right)^{-32/35}\,a_\bh^{5/7}
	~\rm{Myr} .
	\label{eqn:t_gm}
\end{equation}

During the evolution of the BH angular momentum we cap $a_\bh$ at the theoretical limit (in the presence of accreting matter) of $0.998$ \citep[][]{T1974}. Since the total angular momentum of the BH-disc system must be conserved, the disc angular momentum evolves as

\begin{equation}
	\dot{\bm{J}}_\disc\,=\,-\dot{\bm{J}}_\bh\,+\,\dot{\bm{J}}_\infl 
	,
	\label{eqn:dJ_disc/dt}
\end{equation}

\noindent where $\dot{\bm{J}}_\infl$ is the angular momentum change due to the material flowing on to the accretion disc from resolved scales.

As pointed out by \citet{D2013}, in some conditions $R_\warp$ can exceed the total extent of the accretion disc $R_{\rm out}$, i.e. the BH mass is larger than a critical value

\begin{equation}
	M_{\bh,\,\warp}\,\simeq\,10^{7}\,\alpha_{0.1}^{-1/41}\,M_{\disc ,4}^{35/82}\,\left(\frac{\fedd}{\eta_{0.1}}\right)^{-17/82}\,a_\bh^{-25/82}
	\msun .
	\label{eqn:M_warp}
\end{equation}

In this regime, the time-scale for alignment or counter-alignment of the BH and disc angular momenta is drastically reduced and the disc cannot attain a steady warped state. Under these circumstances, we assume that $\bm{J}_\bh$ and $\bm{J}_\disc$ instantaneously reorient themselves towards the axis of the system's total angular momentum, $\bm{J}_{\rm{tot}}$. While $\bm{J}_\bh$ always tends to align itself to $\bm{J}_{\rm{tot}}$, $\bm{J}_\disc$ will end up aligned with $\bm{J}_{\rm{tot}}$ only when \citep[][]{K2005}

\begin{equation}
	\bm{j}_\bh\cdot\bm{j}_\disc\,\geq\,-\frac{J_\disc}{2J_\bh}
	,
	\label{eqn:king_criterion}
\end{equation}

\noindent counter-aligning otherwise.

In order to accurately evolve the spin, we impose a time-step criterion to the sub-grid model, defined as a fraction (fixed at 10 per cent; as long as we properly resolve the investigated time-scales, the presented results are not significantly sensible to the choice of this fraction) of the minimum between the disc consumption time-scale, $\tau_{\rm{drain}} = M_\disc/\dot{M}_{\rm acc-BH}$, and the gravito-magnetic interaction time-scale $\tau_{\rm{gm}}$. Moreover, one of the underlying assumptions of our model is that the disc attains a steady-state warped profile, therefore introducing a further limitation on $\Delta t$, which must be thus greater than the warp propagation time-scale $\tau_{\nu_2}(R_\warp)$ \citep[][]{M2007}:

\begin{equation}
	\Delta t\,=\,\max\left\{\,\tau_{\nu_2}\left(R_\warp\right)\,,\, 0.1\,\min\left\{\,\tau_{\rm{gm}}\,,\,\tau_{\rm{drain}}\,\right\} \,\right\} 
	.
	\label{eqn:dt}
\end{equation}

%%%%%%%%%%%%%%%%%%%%%%%%%%%%%%
%%%%%%%%%%%%%%%%%%%%%%%%%%%%%%
%%%Connecting to simulations
%%%%%%%%%%%%%%%%%%%%%%%%%%%%%%
%%%%%%%%%%%%%%%%%%%%%%%%%%%%%%

\subsection{Connecting the sub-grid model to simulations}\label{sec:connecting_to_sims}

In this section, we describe how we consistently couple our sub-grid model to the hydrodynamics code used for simulations. To this aim, at each time-step, we determine the boundary conditions for the sub-grid evolution from the average properties of the gas surrounding each BH particle. First, we determine the inflow rate $\dot{M}_\infl$ on to the disc as the accretion rate from resolved scales provided by the code.

To prevent the sub-grid disc from becoming self-gravitating, at every time-step we limit $\dot{M}_\infl$ to ensure $M_\disc \le M_\sg$, where $M_\sg$ is computed as the disc mass enclosed within the self-gravitating radius $R = R_\sg$, defined as the radius at which the Toomre parameter \citep[][]{Toomre64} $Q$ is equal to unity:

\begin{equation}
	M_\sg\,\simeq\,2\times 10^{4}\,\alpha_{0.1}^{-1/45}\,M_{\bh,\,6}^{34/45}\,\left(\frac{\fedd}{\eta_{0.1}}\right)^{4/45}
	\msun .
	\label{eqn:M_sg}
\end{equation}

Every time the disc is depleted, we refill it in a self-consistent way by taking the inflow rate from larger scales. The mass of the newly formed disc is instantaneously set to

\begin{equation}
	M_\disc\,=\,\min\left\{ M_{\disc,\,\rm{seed}}\,,\,M_\sg\right\} 
	,
	\label{eqn:M_disc_new}
\end{equation}

\noindent where $M_{\disc,\,\rm{seed}}$ is a user-defined seed-mass parameter for the disc, that we set to $10^5 \msun$. Since the mass inflow from resolved scales could in principle be arbitrarily small, when $\dot{M}_\infl\Delta t\,<M_\disc$ we refill the disc stochastically, with a probability $q= \Delta t\,\dot{M}_\infl/M_\disc$. More specifically, we randomly sample $n\in\left(0,1\right)$ from a uniform distribution: if $n\le q$, the disc is created with mass $M_\disc$; otherwise, its mass is left to zero. We recreate the disc by choosing its angular momentum direction to be along $\bm{j}_{\rm gas}$, and by setting $\fedd = f_{\rm Edd , 0}$, where $f_{\rm Edd,0}$ is a free parameter of the model.

We compute the angular momentum inflow rate on to the disc as

\begin{equation}
    \dot{\bm{J}}_\infl\,=\,\dot{M}_\infl\,\bm{\Lambda}_\infl
    ,
    \label{eqn:J_dot_in}
\end{equation}

\noindent where $\bm{\Lambda}_\infl$ is the angular momentum per unit mass carried by the inflowing material.

Unfortunately, when the angular momentum transfer is not properly resolved, $\bm{\Lambda}_\infl$ is too large to be supported by a self-gravitating disc, hence the gas inflowing from resolved scales cannot circularise and join the accretion disc. In these cases, we reduce $\Lambda_\infl$ to account for any mechanism that would make the gas lose its angular momentum flowing from large scales down to the disc scale. Assuming that the gas circularises at a characteristic radius $R_{\rm{circ}}$, we limit $\Lambda_\infl$ to $J_\disc\left(R_{\rm{circ}}\right) / M_\disc\left(R_{\rm{circ}}\right)$, i.e. to the disc specific angular momentum per unit mass at $R_{\rm{circ}}$. In our model, $R_{\rm circ}$ is specified as a parameter in units of the disc self-gravitating radius $R_\sg$. Our treatment is different from that presented in \citet{F2018}, where they impose $\dot{M}_\infl = 0$ whenever $\Lambda_\infl$ is too large. Their approach could excessively limit the inflow in simulations where the angular momentum transfer is not properly resolved. Moreover, at every time-step, we check if $J_\disc/M_\disc > \Lambda_\isco$. If this condition is not satisfied, then the gas will not be able to settle into circular orbits and will fall on to the BH over time-scales much shorter than $\tau_{\nu_2}(R_\warp)$. In these cases, we instantaneously add the disc mass and angular momentum to the BH and, immediately after, refill the disc as described above.

%%%%%%%%%%%%%%%%%%%%%%%%%%%%%%
%%%%%%%%%%%%%%%%%%%%%%%%%%%%%%
%%%Semi-analytic validation
%%%%%%%%%%%%%%%%%%%%%%%%%%%%%%
%%%%%%%%%%%%%%%%%%%%%%%%%%%%%%

\section{Semi-analytic validation}\label{sec:validation}

\begin{figure}
    \centering
	\includegraphics[width=\columnwidth]{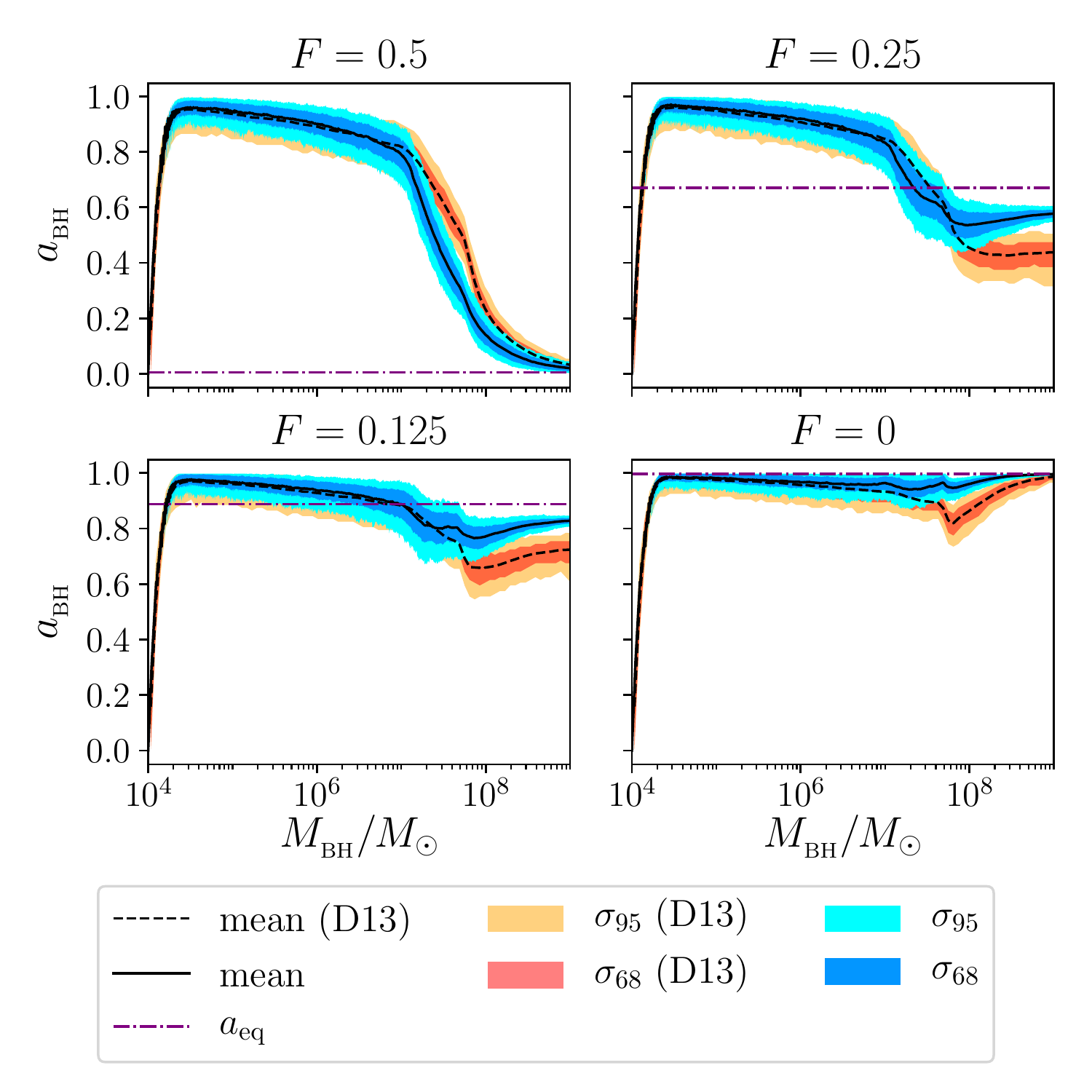}
    \caption{Difference in the evolution of the BH spin parameter $a_\bh$ as the BH mass grows by several orders of magnitude ($10^4$--$10^9 \msun$), assuming $F = 0.5$, 0.25, 0.125, and 0. The black solid line represents the mean over 500 realizations, whereas the shaded areas labelled $\sigma_{68}$ and $\sigma_{95}$ refer, respectively, to the intervals containing 68 and 95 per cent of the data for each mass bin. For an easier comparison, we also report the results obtained by \citeauthor{D2013} (\citeyear{D2013}; D13 in this figure) for the cases we investigated: the black dashed line refers to the mean over 500 realizations, whereas the shaded areas labelled $\sigma_{68}$ (D13) and $\sigma_{95}$ (D13) refer, respectively, to the intervals containing 68 and 95 per cent of the data for each mass bin. The purple, horizontal, dashed-dotted line corresponds to the value of $a_\bh$ analytically computed in the limit of $J_\disc \ll J_\bh$.}
    \label{fig:pallinitest_a_vs_BH}
\end{figure}

To validate our prescriptions, we performed a first test of the BH evolution in a semi-analytically modelled environment, as in \citet{D2013}, wherein an initially non-rotating BH with $M_\bh = 10^4 \msun$ grows in mass through subsequent accretion episodes. At the onset of each episode, we create a new accretion disc with mass $M_\disc = \min \{ M_\sg\, , \,M_{\rm{cloud}} \}$, where $M_{\rm{cloud}} = 10^5 \msun$ is the maximum mass available for each episode. This procedure prevents the disc from becoming too massive for relatively large BH masses, since the cold gas fraction relative to stars is observed to decrease with increasing galaxy mass \citep[e.g.][]{di_Serego_Alighieri_2007,Catinella_2010}. In order to compute the angular momentum of a newly formed disc, we must specify a value for $\fedd$, that determines the magnitude in Equation~\eqref{eqn:J_disc/J_bh}, and extract the direction by a Monte Carlo sampling of the distribution of misalignment angles $\theta$ between the angular momenta of the disc and the larger-scale gas reservoir. We control the degree of anisotropy in the fueling process by introducing a parameter $F$ representing the fraction of discs forming with $\theta > \pi/2$. We set the Eddington ratio to a fiducial average value $\fedd=0.1$. However, we note that varying the choice for $\fedd$ in $[0.01,1]$ does not qualitatively change the long-term BH spin evolution investigated with this model. For this validation test, we also assume that neither accretion nor outflows occur, i.e. $\dot{M}_\infl = \dot{M}_{\rm out} = 0$. 

Our results qualitatively reproduce those obtained by \citet{D2013}, although they exhibit a swifter alignment between the angular momenta, due to the stronger contribution of strongly misaligned configurations in our prescription for the gravito-magnetic torque (see Equation~\ref{eqn:dJ_bh/dt}), whereas the model adopted by \citet{D2013} holds only in the approximation of small misalignments \citep[][]{P2009}. In Figure~\ref{fig:pallinitest_a_vs_BH}, we present the evolution of the BH spin parameter $a_\bh$ through several orders of magnitude in the growth of $M_\bh$ ($10^4$--$10^9 \msun$). Independent of the initial BH-disc configuration, for relatively low BH masses, $a_\bh$ rapidly grows up to its maximum value, because the angular momenta of the BH and disc align themselves on time-scales way shorter than the disc consumption time-scale. When $M_\sg \geq M_{\rm{cloud}}$, the disc is refilled with the same mass at each episode, while $J_\disc / J_\bh$ keeps decreasing as $M_\bh$ grows. Choosing a different value for $M_{\rm cloud}$ would only directly affect the value of $M_\bh$ at which we observe this transition. In this regime, alignment becomes inefficient and the BH can spend more time growing in mass via retrograde accretion, thus significantly reducing $a_\bh$. 

For large BH masses, $\bm{J}_\bh$ aligns itself with the average angular momentum of the gas reservoir, therefore $a_\bh$ evolves towards an equilibrium value $a_{\rm{eq}}$ set by the degree of anisotropy in the fueling process \citep[][]{S2014}. With respect to the prescription adopted by \citet{D2013}, our model predicts a swifter alignment with the reservoir, making prograde accretion events more frequent. Therefore, for large BH masses, $a_\bh$ is biased toward higher values, the disc is recreated with smaller $J_\disc/J_\bh$ at the onset of each accretion event, and we earlier enter the regime where $R_\warp \gtrsim R_{\rm out}$. This results in values for $a_\bh$ at large BH masses that are in better agreement with those computed analytically in the limit for $J_\disc / J_\bh\rightarrow 0$.

\section{Numerical simulations}\label{sec:runs}

In this work, we implement a physically motivated BH spin evolution sub-grid model in the publicly available $N$-body, mesh-less hydrodynamics code {\textsc{gizmo}} \citep[][]{H2015}, descendant of \textsc{gadget2} \citep{S2005} and \textsc{gadget3} \citep{S2008}, although the model is flexible enough that it can be easily transported to other codes. In detail, our prescriptions follow the evolution of a BH-disc system associated to BH/sink particles in the code. Simulations were run on the CINECA cluster MARCONI100, with an typical usage of $\sim 10$ CPU hours per Myr.

\subsection{Numerical setup}\label{sec:setup}

\begin{table*}
	\centering
	\caption{Summary of the parameters assumed for our simulations. The subscript zero refers to quantities evaluated at initialization, i.e. at time $t=0$. The runs' labels are chosen to recall the single parameter changed with respect to the \text{Fiducial} run: \text{Rc} stands for $R_{\rm circ}$; \text{Jd} recalls a change in the initial disc angular momentum through a different choice for $f_{\rm Edd,\,0}$; \text{aBH} stands for $a_{\bh,\,0}$; \text{Md} and \text{MBH} stand for the initial disc and BH masses, respectively. The suffixes \text{UL}, \text{VL}, \text{L}, \text{H}, and \text{VH} refer to an initial configuration where a specific parameter is set to ultra-low, very-low, low, high, and very-high, respectively, relative to the \text{Fiducial} run. In run \text{Fiducial+Feedback}, we couple our model with the biconical-outflow model presented in \citet{Sala2021}.}
	\label{tab:runs}
	\begin{tabular}{ccccccc}
		\hline
		run label & $M_{\bh ,\,0}/{\rm M}_{\sun}$ & $M_{\disc ,\,0}/{\rm M}_{\sun}$ & $f_{\rm{Edd},\,0}$ & $J_{\disc,\,0}/J_{\bh,\,0}$ & $R_{\rm{circ}}/R_\sg$ & $a_{\bh,\,0}$\\
		\hline
		\rm{Fiducial} & $10^7$ & $5\times 10^4$ & $5\times 10^{-3}$ & $2.48$ & $0.5$ & $0.5$\\
		\hline
		\rm{Rc-VL} & $10^7$ & $5\times 10^4$ & $5\times 10^{-3}$ & $2.48$ & $0.1$ & $0.5$\\
		\rm{Rc-L} & $10^7$ & $5\times 10^4$ & $5\times10^{-3}$ & $2.48$ & $0.3$ & $0.5$\\
		\rm{Rc-H} & $10^7$ & $5\times 10^4$ & $5\times 10^{-3}$ & $2.48$ & $0.7$ & $0.5$\\
		\rm{Rc-VH} & $10^7$ & $5\times 10^4$ & $5\times 10^{-3}$ & $2.48$ & $0.9$ & $0.5$\\
		\hline
		\rm{Jd-UL} & $10^7$ & $5\times 10^4$ & $0.1$ & $1.07$ & $0.5$ & $0.5$\\
		\rm{Jd-VL} & $10^7$ & $5\times 10^4$ & $5\times10^{-2}$ & $1.30$ & $0.5$ & $0.5$\\
		\rm{Jd-L} & $10^7$ & $5\times 10^4$ & $10^{-2}$ & $2.04$ & $0.5$ & $0.5$\\
		\hline
		\rm{aBH-L} & $10^7$ & $5\times 10^4$ & $5\times10^{-3}$ & $13.05$ & $0.5$ & $0.1$\\
		\rm{aBH-H} & $10^7$ & $5\times 10^4$ & $5\times 10^{-3}$ & $1.50$ & $0.5$ & $0.8$\\
		\hline
		\rm{Md-L} & $10^7$ & $10^4$ & $5\times 10^{-3}$ & $0.26$ & $0.5$ & $0.5$\\
		\rm{MBH-H} & $5\times10^7$ & $5\times 10^4$ & $5\times 10^{-3}$ & $0.12$ & $0.5$ & $0.5$\\
        \hline
		\rm{Fiducial+Feedback} & $10^7$ & $5\times 10^4$ & $5\times 10^{-3}$ & $2.48$ & $0.5$ & $0.5$\\
		\hline
	\end{tabular}
\end{table*}

We carried out simulations with a single BH particle in an idealised environment resembling a typical galactic nucleus, consisting of a gaseous circumnuclear disc (CND) embedded in a stellar bulge \citep[][]{L2015}. The spherical stellar component of our initial conditions follows a \citet{H1990} density profile,

\begin{equation}
	\rho_{\rm b}\left( r\right)\,=\,\frac{M_{\rm b}}{2\pi}\frac{r_{\rm b}}{r\left(r+r_{\rm b}\right)^3}
	,
	\label{eqn:rho_Hernquist}
\end{equation}

\noindent where $r$ is the radial spherical coordinate, $M_{\rm b}=5\times 10^8$~M$_{\sun{}}$ is the bulge total mass, and $r_{\rm b}=100\pc$ is the bulge scale-radius. The gas particles constitute a rotationally supported exponential disc in vertical hydrostatic equilibrium, with a surface density profile

\begin{equation}
	\Sigma_{\rm CND}\left( R\right)\,=\,\frac{M_{\rm CND}}{2\pi\,R_{\rm CND}^2}\,e^{-R/R_{\rm CND}},
	\label{eqn:Sigma_gas_disc}
\end{equation}

\noindent where $M_{\rm CND} = 10^8 \msun$ is the disc total mass and $R_{\rm CND} = 50 \pc$ is the disc scale-radius. The vertical density profile and velocity field of the disc are calculated by means of the publicly available\footnote{\url{http://www.dfm.uninsubria.it/alupi/software.html}} code {\textsc{gd\_basic}} \citep[][]{L2015} taking into account the global potential of the bulge+disc+BH system. We initialised the gas component assuming an ideal equation of state with $\gamma = 5/3$ and uniform temperature $T = 10^4$~K. In order to reduce the pressure support of the disc and favour gas inflows towards the centre, the simulations are performed with a lower $\gamma = 7/5$, allowing us to mimic a mild radiative cooling without actually employing a dedicated sub-grid model \citep{Dotti_et_al_2009}. We consider two initial BH masses: $10^7$ and $5\times 10^7 \msun$. The mass resolution is $10^3 \msun$ for both star and gas particles, translating into $N_{\rm b} = 5\times 10^5$ stellar particles and $N_{\rm CND} = 10^5$ gas particles. The spatial resolution is determined by the Plummer-equivalent gravitational softening parameter $\varepsilon$, that is fixed at 0.16 and 1~pc for stellar and BH particles, respectively. For gas particles/cells, we employ instead fully adaptive softening, i.e. the gravitational and hydrodynamic resolutions are both defined by the kernel size of each gas element, set to encompass an effective number of neighbours $N_{\rm ngb} = 32$. The minimum gravitational softening/kernel size, that also sets the maximum spatial resolution of the simulation for gas, is set to $\varepsilon_{\rm gas}=0.16$~pc.

In order to couple our sub-grid model with the resolved scales in the hydrodynamics code, we determine the accretion rate on to the BH particle by means of the Bondi--Hoyle--Lyttleton \citep{Bondi_Hoyle_1944,Bondi_1952,Hoyle_Lyttleton_1939} formula,\footnote{We note, however, that any other prescription is equally valid.} as implemented by \citet{SDMH2005}:

\begin{equation}
    \dot{M}_\infl=\frac{4\pi \alpha_{\rm acc}G^2 M^2_{\rm BH} \rho}{(c_{\rm s}^2+v^2)^{3/2}}
    ,
\end{equation}

\noindent where $\rho$ is the density of the surrounding gas and $v$ is the gas-BH relative velocity, both determined via mass-weighting over the nearest $N_{\rm ngb,BH}$ neighbour particles of the BH. In our implementation, we set $N_{\rm ngb,BH} = 3 N_{\rm ngb}$. Finally, $\alpha_{\rm acc}$ is a dimensionless parameter, typically employed to correct for the interstellar medium's dense gas not properly resolved in $\sim$kpc-scale simulations \citep{DiMatteo05,Booth09}, which we set equal to one. In order to prevent the accretion properties from being un-physically affected by gas particles at large distances, we set a maximum accretion radius of 10~pc. Therefore, the BH particle kernel is defined as the region encompassing  $N_{\rm ngb,BH}$ neighbours, unless limited in size by the specified maximum accretion radius.

To compute the angular momentum inflow rate on to the disc, $\dot{\bm{J}}_\infl$, we assume that the angular momentum per unit mass carried by the inflowing material, $\bm{\Lambda}_\infl$, is equal to $\bm{J}_{\rm{gas}} / M_{\rm{gas}}$, where $M_{\rm{gas}}$ is the total mass of the gas particles in the BH particle kernel, and $\bm{J}_{\rm{gas}}$ is their total angular momentum.The direction of $\bm{J}_{\rm{gas}}$ is also used to initialise $\bm{j}_\disc$. We performed simulations initialising the BH-disc angular momenta misalignment angle to $\theta_{\bh-\disc} = \arccos\left(\bm{j}_\bh\cdot\bm{j}_\disc\right) = 5\pi/6$, and varying the initial BH and disc masses and angular momentum ratio, as well as the parameter $R_{\rm circ}/R_\sg$. As fiducial parameters for our model, adopted to initialise our \text{Fiducial} run, we took: $M_{\bh, 0} = 10^7 \msun$; $M_{\disc, 0} = 5\times10^4 \msun$;  $f_{\rm Edd, 0} = 5\times10^{-3}$;\footnote{The parameter $f_{\rm Edd,\,0}$ is only needed to initialise the disc properties, and $\fedd$ self-adjusts according to $\dot{M}_\infl$ during the first time-steps.} $R_{\rm circ}/R_{\rm sg} = 0.5$; and $a_{\bh, 0} = 0.5$. Details on the choice of parameters for each performed simulation are reported in Table~\ref{tab:runs}.

\subsection{Results}

\begin{figure}
    \centering
	\includegraphics[width=\columnwidth]{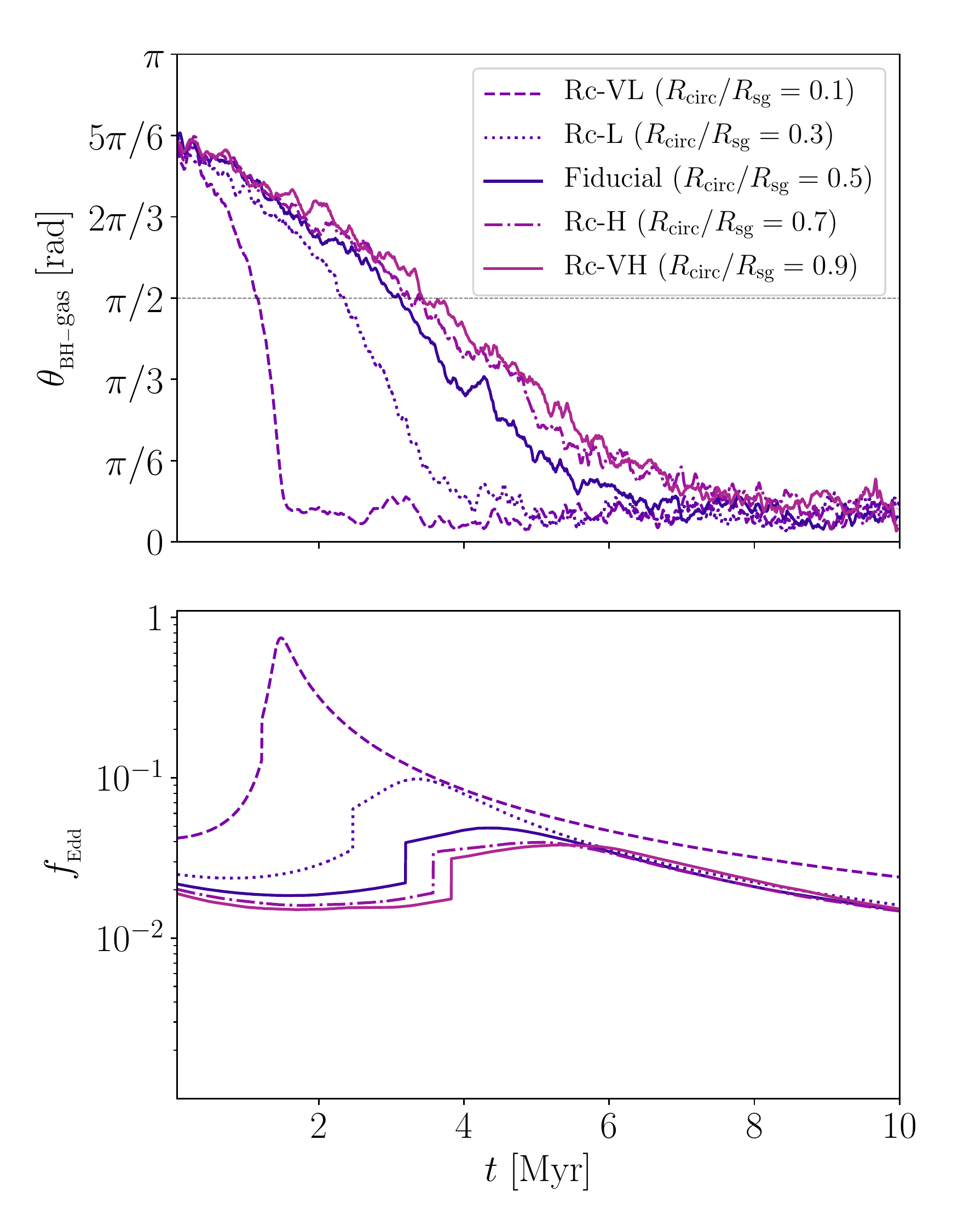}
 	\caption{Time evolution of the misalignment angle $\theta_{\bh-\rm gas}$ (top panel) and the Eddington ratio $\fedd$ (bottom panel), in runs \text{Rc-VL}, \text{Rc-L}, \text{Fiducial}, \text{Rc-H}, and \text{Rc-VH}. These runs share the same set of initial parameters, but for $R_{\rm circ}/R_\sg$. A smaller $R_{\rm circ}$ implies a smaller angular momentum inflow, resulting in a more compact disc with higher accretion rate and in a faster BH spin evolution.}
    \label{fig:figure2}
\end{figure}

In Figure~\ref{fig:figure2}, we show the effect of varying the parameter $R_{\rm circ}/R_\sg$, which controls the angular momentum inflow on to the disc, on the misalignment angle $\theta_{\bh-\rm gas} = \arccos\left(\bm{j}_\bh\cdot\bm{j}_{\rm gas}\right)$ between the BH spin and the average angular momentum of the resolved gas reservoir (top panel), and on the Eddington ratio $\fedd$ (bottom panel). We compare runs \text{Rc-VL}, \text{Rc-L}, \text{Fiducial}, \text{Rc-H}, and \text{Rc-VH}, for which we assume the same BH-disc initial configuration (see Table~\ref{tab:runs} for details). A smaller circularisation radius corresponds to a smaller upper limit for $\Lambda_\infl$, in turn related to a more compact disc (at fixed $M_\disc$) and therefore to a larger accretion rate on to the BH (i.e. a larger $\fedd$). This results in a swifter BH spin evolution for smaller values of $R_{\rm circ}/R_\sg$ and the difference in the evolution is more evident the smaller $R_{\rm circ}/R_\sg$ is.

\begin{figure}
    \centering
	\includegraphics[width=\columnwidth]{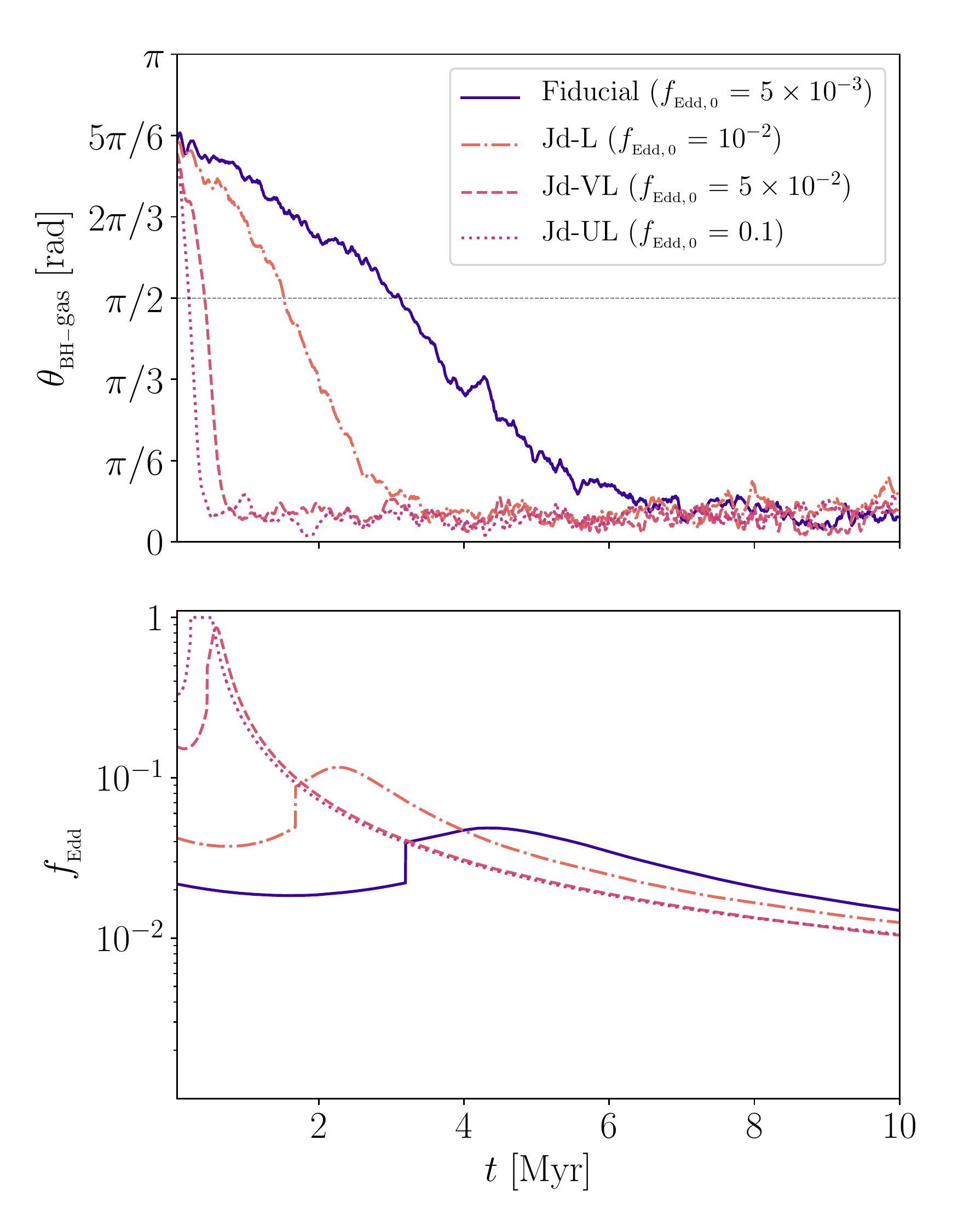}
	\caption{Time evolution of the misalignment angle $\theta_{\bh-\rm gas}$ (top panel) and the Eddington ratio $\fedd$ (bottom panel), in runs \text{Fiducial}, \text{Jd-L}, \text{Jd-UL}, and \text{Jd-VL}. These runs share the same set of initial parameters, but for $f_{\rm Edd,\,0}$. The smaller the initial Eddington ratio $f_{\rm Edd,\,0}$ (i.e. larger initial total angular momentum of the disc), the slower is the BH spin evolution, because of a reduced accretion rate on to the BH from the less compact disc.}
    \label{fig:figure3}
\end{figure}

In Figure~\ref{fig:figure3}, we show the impact of changing the initial Eddington ratio $f_{\rm Edd,\,0}$ on the time evolution of $\theta_{\bh - \rm gas}$, by comparing runs \text{Jd-UL}, \text{Jd-VL}, \text{Jd-L}, and \text{Fiducial}, with $f_{\rm Edd , 0} = 0.1$, $5 \times 10^{-2}$, $10^{-2}$, and $5 \times 10^{-3}$, respectively. Configurations with larger $f_{\rm Edd , 0}$ correspond to lower initial $J_\disc/J_\bh$, resulting in more compact and denser discs with higher accretion rates, and leading to a faster evolution of the BH spin.

\begin{figure*}
    \centering
	\includegraphics[width=\textwidth]{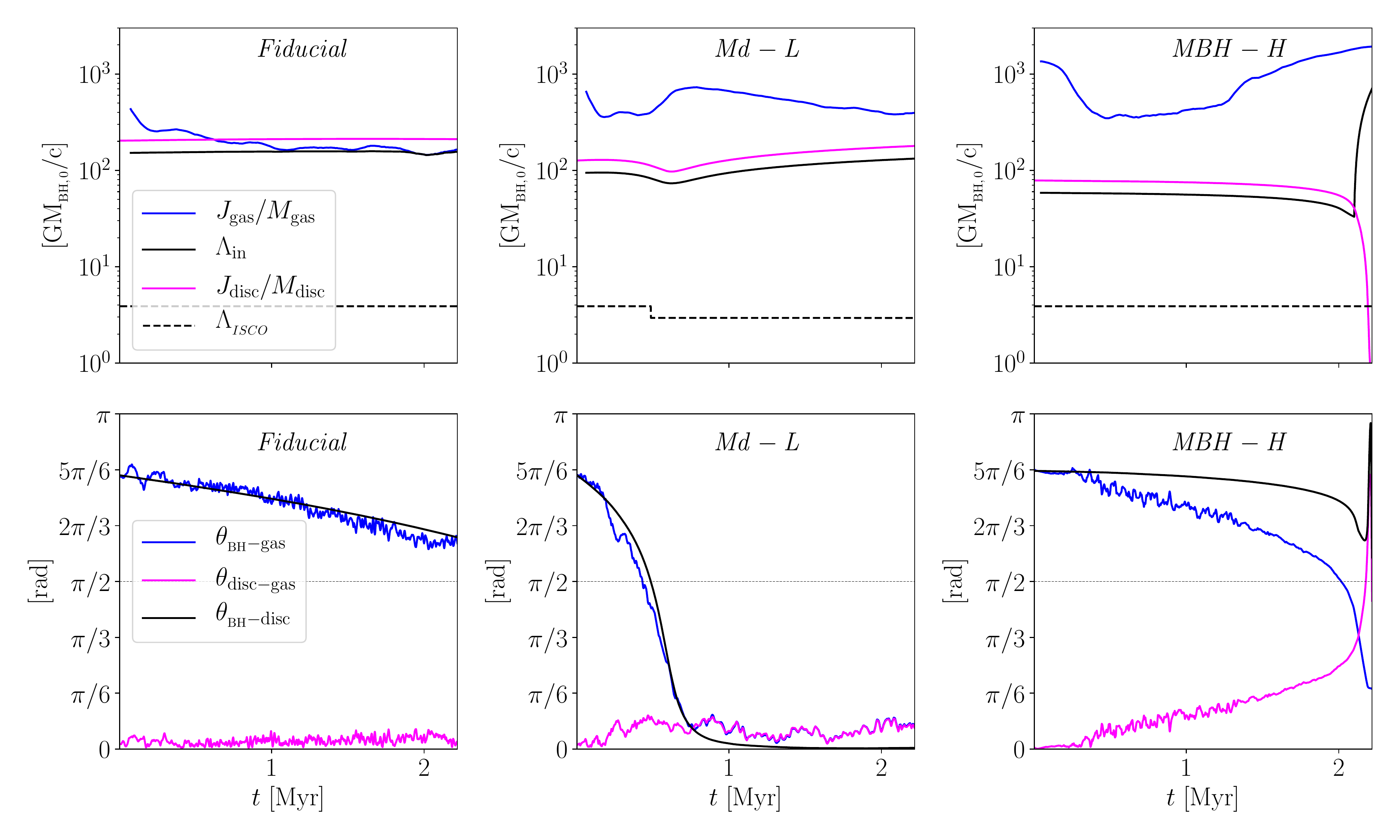}
	\caption{We compare the first few Myr of runs \text{Fiducial} (left-hand panels), \text{Md-L} (central), and \text{MBH-H} (right-hand). With respect to our \text{Fiducial} run, run \text{Md-L} is initialised with a less massive disc ($M_{\disc,\,0} = 10^4 \msun$), whereas run \text{MBH-H} presents a BH with a larger initial mass ($M_{\bh,\,0} = 5 \times 10^7 \msun$). In the top panels, we compare the evolution of $J_\disc/M_\disc$ (magenta solid lines), $\Lambda_\infl$ (black solid lines), and $J_{\rm gas}/M_{\rm gas}$ (blue solid lines). The black dashed lines represents $\Lambda_\isco$. In run \text{Md-L}, $\Lambda_\isco$ exhibits a clear jump at $\sim$0.5~Myr, when the BH and disc switch from co-rotating to counter-rotating, because the ISCO radius differs in the two scenarios. This behaviour of $\Lambda_\isco$ is seen in every run in which there is a switch between counter-rotating and co-rotating configurations (e.g. in the \text{Fiducial} run, but at later times than those presented in this figure). In the bottom panels, we show the evolution of $\theta_{\bh - \disc}$ (black line), $\theta_{\bh - \rm gas}$ (blue line), and $\theta_{\disc - \rm gas}$ (magenta line). The more compact disc of run \text{Md-L} results in the BH and disc angular momenta ending up aligned over $\sim$1~Myr. In run \text{MBH-H}, after an initial slow evolution towards alignment, the BH and disc angular momenta rapidly counter-align.}
    \label{fig:figure4}
\end{figure*}

In runs \text{Md-L} and \text{MBH-H}, we changed the initial disc and BH mass, respectively, with respect to our \text{Fiducial} run. In run \text{Md-L}, we assume an initially less massive disc, with $M_{\disc,\,0} = 10^4 \msun$. In run \text{MBH-H}, we chose a heavier BH seed, with $M_{\bh,\,0} = 5 \times 10^7 \msun$. The different time evolution of $\theta_{\bh-\disc}$, $\theta_{\bh-\rm gas}$, and $\theta_{\disc-\rm gas}= \arccos\left(\bm{j}_\disc\cdot\bm{j}_{\rm gas}\right)$, i.e. the misalignment angles between the angular momenta of the BH, disc, and gas reservoir in these runs, is presented in the lower panels of Figure~\ref{fig:figure4}. The upper panels show the time evolution of $J_{\rm gas}/M_{\rm gas}$ (blue line) and $\Lambda_\infl$ (black solid line), until $J_\disc/M_\disc$ (magenta line) in run \text{MBH-H} drops below $\Lambda_\isco$ (dashed black line). All these runs share the same initial accretion rate $f_{\rm Edd,\,0} = 5 \times 10^{-3}$. This translates in the $10^4 \msun$ disc of run \text{Md-L} being initially more compact (less angular momentum, with equal $\fedd$) with respect to runs \text{Fiducial} and \text{MBH-H}, hence in a faster BH evolution. In run \text{MBH-H}, the circularisation radius is closer to $R_\warp$ than in the other runs, because of the more massive BH. The major contribution to the disc total angular momentum comes from regions closer to the BH, where we expect the disc to significantly modify its angular momentum direction. Both $\bm{J}_\disc$ and $\bm{J}_\bh$ tend to align/counter-align with $\bm{J}_{\rm tot}$. Therefore, a lower $J_\disc/J_\bh$, i.e. a reduced contribution of $\bm{J}_\disc$ to $\bm{J}_{\rm tot}$, results in a more evident evolution of $\bm{J}_\disc$.

In run \text{Md-L}, In the \text{Fiducial} run, this effect is negligible, since the disc is more extended and less compact, whereas in run \text{MBH-H}, the evolution of $\bm{J}_\disc$ is much more pronounced. Moreover, $J_\disc/J_\bh$ is small enough to expect the BH-disc system to evolve towards a configuration with counter-aligned angular momenta. Indeed, the angle $\theta_{\bh - \disc}$ (black line in the bottom panels of Figure~\ref{fig:figure4}) eventually settles to $\sim\pi$, whereas the angle $\theta_{\bh - \rm gas}$ (blue line in the bottom panel) approaches zero. Before $J_\disc/M_\disc \lesssim \Lambda_\isco$, the BH and disc angular momenta counter-align, with the BH spin close to being aligned with the average angular momentum direction of the reservoir, $\bm{j}_{\rm gas}$.  Therefore, $\bm{J}_\disc$ is almost counter-aligned with respect to $\bm{j}_{\rm gas}$, and the coherent inflow of angular momentum directed as $\bm{j}_{\rm gas}$ acts to reduce $J_\disc/M_\disc$, which rapidly drops below the threshold of $\Lambda_\isco$. In the evolution of $\bm{J}_\disc$, our model assumes that the inflowing counter-rotating gas and the disc shock instantaneously, settling to a more compact configuration with less angular momentum. This behaviour is not necessarily physical,  but it is the best that can be done with a model that does not fully evolve self-consistently the whole angular momentum profile of the accretion disc. Such an improvement would significantly slow down any large scale simulation, and therefore is not considered in this study.

\begin{figure}
    \centering
	\includegraphics[width=\columnwidth]{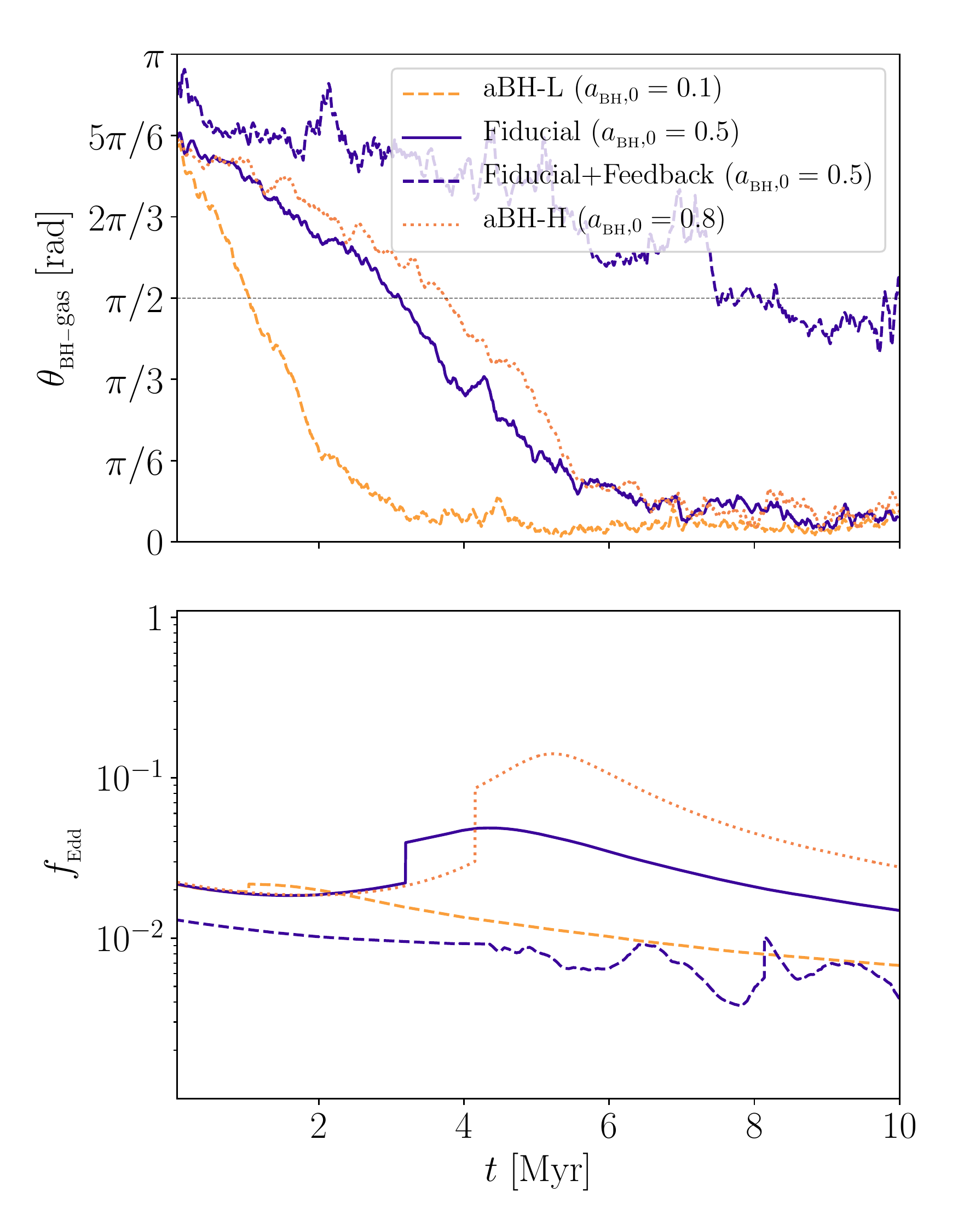}
	\caption{Time evolution of the misalignment angle $\theta_{\bh-\rm gas}$ (top panel) and the Eddington ratio $\fedd$ (bottom panel), in runs \text{aBH-L}, \text{aBH-H}, \text{Fiducial}, and \text{Fiducial+Feedback}. These runs share the same set of initial parameters, but for $a_{\bh,\,0}$. Runs \text{Fiducial+Feedback} and \text{Fiducial} share the same $a_{\bh,\,0}$, but in the former we couple our model with the biconical-outflow model presented in \citet{Sala2021}. A higher initial $a_\bh$ results in a delayed alignment between the BH and disc angular momenta, and in a higher accretion rate, after a transition period of a few Myr.}
    \label{fig:figure5}
\end{figure}

In Figure~\ref{fig:figure5}, we present the effect of changing the initial BH spin parameter $a_{\bh,\,0}$ on the evolution of $\theta_{\bh - \rm gas}$, by comparing runs \text{aBH-L}, \text{Fiducial}, and \text{aBH-H}, where we set $a_{\bh,\,0}$ equal to 0.1, 0.5, and 0.8, respectively. A more rapidly spinning BH offers more resistance to changes in its angular momentum direction, resulting in a slower spin evolution (see top panel of Figure~\ref{fig:figure5}) and, after a transition period of a few Myr, higher accretion rates (bottom panel). In run \text{aBH-H}, accretion switches from retrograde to prograde at later times, allowing for $\fedd$ to peak at larger values, up to an order of magnitude with respect to run \text{aBH-L}.

Finally, in our \text{Fiducial+Feedback} run, we coupled the spin evolution model to the new accretion/feedback model by \citet{Sala2021}, assuming that biconical outflows are launched along the BH spin direction, and the outflow rate $\dot{M}_{\rm out}$ is self-consistently determined by the actual BH accretion rate provided by the accretion disc. The evolution of $\theta_{\rm BH-gas}$ with time  for this last run is shown in the top panel of Figure~\ref{fig:figure5} as a blue dashed line, to facilitate the comparison with the no-feedback sibling run (\text{Fiducial}). While the initial evolution (up to $\sim 1$ Myr) is similar, with a slightly slower alignment in the simulation with feedback due to the reduced mass accretion rate (accommodating for the generated outflow), the late evolution differs significantly. This is due to the increased relevance of the biconical feedback when the spin direction approaches the large-scale gaseous disc mid-plane and the feedback therefore impinges on to the dense gas distribution, strongly affecting its dynamics and further reducing the inflow rate on to the BH-disc system. These effects can be clearly seen in the bottom panel of Figure~\ref{fig:figure5}, where the accretion rate in \text{Fiducial+Feedback} is only a factor of two lower than in \text{Fiducial} during the first $\sim$Myr of the run, but then the difference increases up to a factor of $\sim$4--5.

\section{Conclusions}\label{sec:conclusions}

We have introduced a novel sub-grid model for the evolution of the BH mass and spin and the surrounding accretion disc. Our model is a direct descendant of the model presented in \citet{F2018}, with some modifications concerning the limitation on the angular momentum inflow through $R_{\rm circ}$ and the stochastic refill of the disc mass in the case of small mass inflow rates, as discussed in Section~\ref{sec:connecting_to_sims}. Moreover, we do not impose any lower limit to $\fedd$, because the BH spin evolution can be neglected for very small accretion rates.
In addition, the newly implemented model for the unresolved accretion disc accounts for the presence of self-consistently computed outflows \citep[see][for a detailed description of the implementation]{Sala2021}.

In the future, we are planning to further improve our model, in order to increase its accuracy and account for additional processes that can affect the BH spin. In particular, we will \textit{(i)} employ a more sophisticated viscosity prescription, based on the numerical solutions for different warps \citep[see, e.g.][]{O2013,Tremaine_Davis_2014}, \textit{(ii)} account for the additional angular momentum coupling in BH binaries in gaseous environments \citep[e.g.][]{Gerosa_et_al_2020} and the spin change after BH mergers, and \textit{(iii)} consider a more appropriate coupling between resolved and sub-grid scales when the accretion disc is counter-aligned relative to the inflowing gas, that would relax the instantaneous shock assumption employed in this study.

We first assessed the validity of our model in a semi-analytically modelled environment, qualitatively reproducing the results of \citet{D2013}. A crucial quantitative difference, however, is represented by the swifter alignment between the angular momenta in our model, that results from its validity also for strongly misaligned configurations. 

We then interfaced our model with the publicly available code {\textsc{gizmo}}. We itemize the findings of our $N$-body, hydrodynamic tests below:

\begin{itemize}
    
\item The radius at which the gas circularises, $R_{\rm circ}$, has profound effects on the BH evolution, with a swifter spin evolution and a higher accretion rate for smaller $R_{\rm circ}$.
    
\item The initial total angular momentum of the disc, $J_{\rm disc}$, also plays an important role in the BH spin evolution, that becomes slower for larger values of $J_{\rm disc}$ (or, equivalently, for smaller values of the initial Eddington ratio).
    
\item Decreasing the initial accretion disc mass produces a disc with less angular momentum. The larger growth rate for $M_\disc$ than for $J_\disc$ translates into a larger $\fedd$ at later times and, therefore, in a faster BH evolution.
    
\item An initial larger (smaller) BH spin results in a slower (faster) evolution, due to increased (reduced) resistance to changes in its own angular momentum.
    
\item The inclusion of BH feedback has the effect of altering the dynamics of the CND and reducing the gas inflow, in particular when the spin evolution leads to the feedback cone to cross the large-scale gas distribution. This, as a consequence, further slows down the evolution of the spin direction, when its relative angle respect to the CND angular momentum is about $\pi/2$.
    
\end{itemize}

In conclusion, our model is able to accurately follow the BH mass and spin evolution in hydrodynamic simulations, also when coupled with a sub-grid prescription for BH accretion and feedback, and can be easily applied to simulations on different scales, from galaxy mergers to cosmological simulations. In future works, we will employ it to investigate the evolution of BH spin in a cosmological environment, enabling us to make prediction for the Laser Interferometer Space Antenna \citep[][]{Amaro-Seoane2017,Barack_et_al_2019} and pulsar timing arrays.

\section*{Acknowledgements}
We thank the anonymous referee for useful comments that helped to improve the manuscript. We acknowledge the CINECA award under the ISCRA initiative for the availability of high-performance computing resources and support (projects numbers HP10CFXS9S and HP10CRRSO8). AL acknowledges support from the European Research Council Advanced Grant N. 740120 ‘INTERSTELLAR’. This work reflects only the authors’ view and the European Research Commission is not responsible for information it contains. LS acknowledges support from ‘BiD4BEST’ - European Innovative Training Network (ITN) funded by The Marie Soko{\l}owska-Curie Actions (860744) in Horizon 2020

\section*{Data Availability Statement}
The data underlying this article will be shared on reasonable request to the corresponding author.

%%%%%%%%%%%%%%%%%%%%%%%%%%%%%%%%%%%%%%%%%%%%%%%%%%



\bibliographystyle{mnras}
\bibliography{paper} % if your bibtex file is called paper.bib

%\begin{thebibliography}{99}
%\bibitem[\protect\citeauthoryear{Author}{2012}]{Author2012} Author A.~N., 2013, Journal of Improbable Astronomy, 1, 1
%\bibitem[\protect\citeauthoryear{Others}{2013}]{Others2013} Others S., 2012, Journal of Interesting Stuff, 17, 198
%\end{thebibliography}

%%%%%%%%%%%%%%%%%%%%%%%%%%%%%%%%%%%%%%%%%%%%%%%%%%

%%%%%%%%%%%%%%%%% APPENDICES %%%%%%%%%%%%%%%%%%%%%

\appendix

%If you want to present additional material which would interrupt the flow of the main paper, it can be placed in an Appendix which appears after the list of references.

%%%%%%%%%%%%%%%%%%%%%%%%%%%%%%%%%%%%%%%%%%%%%%%%%%

% Don't change these lines
\bsp	% typesetting comment
\label{lastpage}
\end{document}